# Telecom light-emitting diodes based on nanoconfined self-assembled silicon-based color centers

Andreas Salomon[1], Johannes Aberl[1], Enrique Prado Navarrete[1], Merve Karaman[1], Ádám Gali[2-4], Thomas Fromherz[1], Moritz Brehm[1]

[1] *Institute of Semiconductor and Solid State Physics, Johannes Kepler University, Altenberger Straße 69, Linz 4040, Austria*

[2] *HUN-REN Wigner Research Centre for Physics, P.O. Box 49, H-1525, Budapest, Hungary*

[3] *Department of Atomic Physics, Institute of Physics, Budapest University of Technology and Economics, Műegyetem rakpart 3., Budapest H-1111, Hungary*

[4] *MTA-WFK Lendület "Momentum" Semiconductor Nanostructures Research Group, P.O. Box 49, H-1525, Budapest, Hungary*

**Abstract**

**Silicon color centers (SiCCs) have recently emerged as potential building blocks for light emitters in Si photonics, quantum emitters with spin storage capabilities, and Si-based quantum repeaters. We have recently developed a non-invasive method to engineer carbon-related SiCCs confined to ultra-thin nanolayers within a pristine crystalline environment, which is of utmost importance for the photostability of SiCCs. Here, we demonstrate embedding these C-doping-based SiCCs into the only 9 nm wide intrinsic region of a p-i-n diode using the epitaxial self-assembly of color centers. We report electrically-pumped light emission with an exponential increase in the intensity as a function of the driving current until saturation. We associate this property with the shift of quasi-Fermi-level position upon electrical driving, which simultaneously improves the spectral homogeneity of the engineered SiCCs. Our study demonstrates the electrical control and driving of near-infrared emitters in high-quality silicon diodes, an essential milestone for advancing classical and quantum optoelectronics.**

## Introduction

Silicon color centers (SiCCs) have been studied since the 70s as detrimental side effects caused by radiation damage in the fabrication of microelectronic devices [1]. However, recently, they experienced a renaissance due to their peculiar light emission properties. On the one hand, SiCCs such as the so-called G-center, a carbon-based point defect composed of two substitutional carbon atoms and one interstitial Si atom [1,2], have been shown to provide optical gain [3] for potential lasing applications in the telecom spectral range. In that respect, efficient Si-based light emission for optical interconnects still remains a missing piece and represents the "holy grail" of silicon photonics [4,5]. On the other hand, several types of isolated SiCCs, including single G-centers, W-centers, or T-centers have recently exhibited promising quantum properties such as high-purity single photon emission and spin manipulation capabilities [6-10]. In particular, the G-center established itself as a promising candidate for the realization of spin-photon interfaces [2,11] that can particularly benefit from the CC's compatibility with Si and SOI integration technology. However, for both, SiCC-based lasers and quantum-emitters, embedding them into a p-(i)-n diode is of key importance. Useful integrated lasers must be driven electrically [12], and diodes operated in reverse bias conditions are used to clear the charge environment and limit blinking and spectral diffusion of quantum emitters [13]. Conventionally, however, CC creation is enabled by ion implantation, which causes collateral damage in the crystal lattice and a broad emitter distribution due to Gaussian implantation profiles [14]. Compared to optically excited luminescence samples, the effects of ion implementation are even worse in vertical diode structures, as the ion implantation needs to be performed through the top contact layer to avoid a parasitic interface in the vicinity of the respective emitter(s) [16,17].

Here, we demonstrate that carbon-based SiCC light-emitting diodes (LEDs) can be fabricated without the detrimental side effects of ion implantation by solely using epitaxy for the formation of the p-n diode as well as the CCs. Importantly, the CCs can be formed deterministically in nanometer-thin layers of C-doped Si that can be grown exactly in the ultra-thin intrinsic region of the p-i-n structure of the diode. Ultra-low growth temperatures (≤ 300°C) and excellent growth pressures are needed to (i) enable the *in-situ* formation of SiCCs with excellent optical properties, (ii) avoid SiCC dissociation during top contact growth maintaining a high crystalline quality, and (iii) enable an optimal device functionality with an emission that is not inferior as compared to optically pumped samples. We show that despite the low required growth temperatures, epitaxial growth of diodes is feasible, and we demonstrate pronounced CC emission at low current densities < 1 $A/cm^2$ from a SiCC layer as thin as 9 nm located at the p-n junction of the diode. Thus, we present a proof-of-concept fabrication method that demonstrates that electrically-driven SiCCs can be deterministically confined down to layer thicknesses less than 10 nm, opening up the path for potential applications as light sources in Si-based photonics and scalable integrated quantum technology. Furthermore, we show that the present electric field leads to an improved spectral homogeneity of the engineered SiCCs.

**Methods**

The samples for electroluminescence (EL) investigations were grown on Czochralski, p-type, B-doped 4-inch Si(001) substrates (0.001-0.005 Ωcm) using solid source molecular beam epitaxy in a Riber SIVA-45 chamber. The doped substrate provides the p-region of the p-i-n diode. The base pressure in the chamber was $5\times10^{-11}$ mbar, and the maximum pressure during the growth was $3\times10^{-10}$ mbar. The growth protocol can be found in Fig. S1 of the supplementary material. The samples for photoluminescence (PL) characterization were grown on FZ Si(001) wafers with resistivities > 5000 Ωcm. All substrates were cleaned using conventional Si cleaning, see Ref. [18]. After loading the samples into the MBE chamber, they were initially degassed in an ultra-high vacuum at 700°C for 15 min. Hereafter, the samples were kept at 450°C for 30 min. Initially, a p-type, B-doped Si buffer layer with a doping concentration of $5{\cdot}10^{18}$ $cm^{-3}$ and a thickness of 50 nm was grown at a growth temperature of 500°C. After that, the substrate temperature was ramped down to 200°C, the growth temperature of the C-doped Si layer. At a C-doping concentration of $3.8{\cdot}10^{19}$ $cm^{-3}$, a 9 nm thick Si:C layer was deposited. Thereafter, the substrate temperature was ramped to 300°C, at which the n-type top contact of the diode was grown. It consisted of a 50 nm thick Si:Sb layer with a doping concentration of $5{\cdot}10^{18}$ $cm^{-3}$, followed by a 150 nm thick Si:Sb layer with a doping concentration of $2{\cdot}10^{19}$ $cm^{-3}$, leading to symmetric doping concentrations around the intrinsic Si:C layer hosting the CCs. As a reference, an all Si diode was grown, for which the Si:C layer was replaced by a 9 nm thick intrinsic Si layer, deposited at 310°C. For the complementary PL investigations of the C-doping concentration dependence on the luminescence properties of the SiCCs, several additional reference samples were fabricated. Here, after a 75.5 nm thick intrinsic Si buffer, Si:C layers with a common thickness of 9 nm (similar to the diode structures described above), but with variable C concentration ranging from $2.2{\cdot}10^{17}$ $cm^{-3}$ to $5.0{\cdot}10^{20}$ $cm^{-3}$, were grown at a temperature of 200°C. The C deposition rates were calibrated using secondary-ion mass spectrometry (SIMS) experiments of calibration layers. Finally, the samples were capped with intrinsic Si of 105 nm thickness deposited at 300°C.

We processed the MBE-grown p-i-n structures (Fig. 1) into square mesa diodes with side lengths of 400 µm using standard Si processing techniques in a cleanroom environment. To enable EL emission perpendicular to the diode surface, the electrical metal contact (3 nm of Ti and 400 nm of Au) to the top n-type layer was implemented as a ring and grid contact, see inset of Fig. 1(c). As the bottom contact, at the substrate's backside, the same layer structure of Ti-Au was used, followed by conductive silver paste bonding to a chip carrier. No sidewall passivation was applied. Gold wire bonding was performed to drive the current to the device's top. The CC diode samples were driven by a Keithley 2601-B Pulse source measure unit (SMU). Current-voltage (I-V) curves for a current up to 300 mA were recorded in continuous mode by increasing voltage steps, while for higher currents, the SMU was operated in pulsed mode with 50 µs current pulses to avoid excessive device heating and thermal overloading of the bonding wires.

Voltage measurements were done at the device contacts (4-point method) to exclude the voltage drop caused by the connection strings. The EL data was recorded in a sample temperature range from 5 K to 100 K, with the emitted light collected by an infinity-corrected 10× Olympus objective with a numerical aperture of 0.26. The collected signal was coupled free-space into a (Teledyne Princeton Instruments) SpectraPro HRS-750 Czerny-Turner type spectrometer with switchable gratings and a connected liquid nitrogen-cooled linear 1024 pixel InGaAs photodiode array with an effective cut-off energy of 0.775 eV (1600 nm). Additionally, we performed micro-photoluminescence (µ-PL) measurements at low sample temperatures. For excitation, we used a continuous-wave (cw) diode-pumped solid-state (DPSS) laser emitting at 473 nm and a laser power of max. 6 mW (measured below the cryostat window). The laser was focused, and the luminescence signal was collected via an infinity-corrected microscope objective with 0.26 numerical aperture (NA) for the ensemble measurements. The µ-PL spectra were recorded via a 500 mm focal-length Czerny-Turner spectrometer with three interchangeable ruled gratings (100, 300, and 900 l/mm) connected to a liquid-nitrogen (LN2) cooled 1024 pixel InGaAs line detector.

**Results**

The common fabrication of CCs, in general, and SiCCs, in particular, involves either single- or multi-step ion implantation. For example, for the fabrication of G-centers on Si or SOI, implantation energies for C ions >10 keV are used [6-8], which, on the one hand, provide the targeted stopping range for color center emission but, on the other hand, lead to a gaussian distribution profile for SiCC creation that spreads throughout hundreds of nm (see Fig. 1(a)) [14,18]. This spread leads to intrinsic constraints in photonic device designs, i.e., photonic mode – emitter coupling and variations among device ensembles. Implantation at lower energies could, in principle, limit the emitter spread but leads to the creation of emitters close to the surface. Surfaces and interfaces are always troublesome regarding their interaction with the (quantum)-emitters. Compared to SiCCs that are excited optically, the emitter spread is typically even worse for electrically pumped devices since the implantation occurs through thick top-contact layers [14,22-24], which calls for higher implantation energies and thus leads to an even larger emitter spread see Fig. 1(a). Note that the emitter spread worsens further if channeling effects are present during the ion implantation. Lateral diodes that underwent ion implantation are still subjected to the same vertical emitter distribution problem as optically-pumped samples [25].

Here, we seek an alternative fabrication scheme for SiCCs within a light-emitting diode, using only epitaxy. We emphasize that the top contact formation after the emitter creation asks for cautiousness regarding the employed growth conditions since SiCCs are fragile with respect to thermal annealing, i.e., G'-centers, and W-centers annihilate already at annealing temperatures ranging from 200°C to 300°C [1,18]. The thermal load to which the SiCCs are exposed is also a concern for the here presented CCs that result from epitaxial self-assembly during the growth. Figure 1(b) depicts the sample structure of the SiCCs confined to a 9 nm thick layer and embedded in the ultra-thin i-region of the p-i-n junction of the semiconductor diode. Based on previous work [18], we have grown the SiCC layer at 200°C and using a relatively high C-concentration of $3.8{\cdot}10^{19}$ $cm^{-3}$, for which a strong luminescence response can be obtained. We note that the resulting SiCC is predominantly not the traditional G-Center that is known to have a zero-phonon line (ZPL) at ~1278 nm. Instead, here we find a derivate of the conventional G-center that we labeled G'-center before. *Ab initio* calculations of various atomistic configurations strongly indicate that the G'-center consists of a conventional G-center, i.e., two substitutional C atoms combined with an interstitial Si atom ($Si_i$), that is accompanied by an additional substitutional C-atom, see Fig. 1(c) [18]. Such a defect configuration leads to the observed wavelength shift of the ZPL to ~1300 nm while conserving the phonon spectrum properties of the G-center and its favorable spin properties [18]. Such an atomic configuration is likely since our MBE C-effusion cell is no pure source of atomic C but, in addition, emits molecules with three C-atoms that might be built in as a unit into the Si crystal as a consequence of the low growth temperature of 200°C, facilitating the dominant formation of G'-centers.

In this previous work, which relies on PL studies of the SiCC emission only, we found that an overgrowth temperature of 300°C provides a good compromise between preservation of the SiCCs, i.e., the absence of significant thermal SiCC annihilation, and an excellent epitaxial quality of the Si capping layer. We note that we intentionally grew the p-type contact at the bottom and the n-type contact at the top of the layer sequence since dopant activation is possible at lower temperatures for Sb than for B [20,21]. Thus, the higher temperature needed for B-doping activation is employed before the growth of the SiCC layer. Hence, no further annealing steps after the diode growth have been required to activate the dopants, facilitating the preservation of the optical properties of the created SiCCs. Figure 1(d) depicts temperature-dependent I-V curves, showing clear diode characteristics and demonstrating that the low growth temperature used to grow the Si:Sb top contact layer is not detrimental to diode functionality. The insert depicts a photograph of the processed device in which the Ti-Au top-contact grid and the bonding wires are visible.

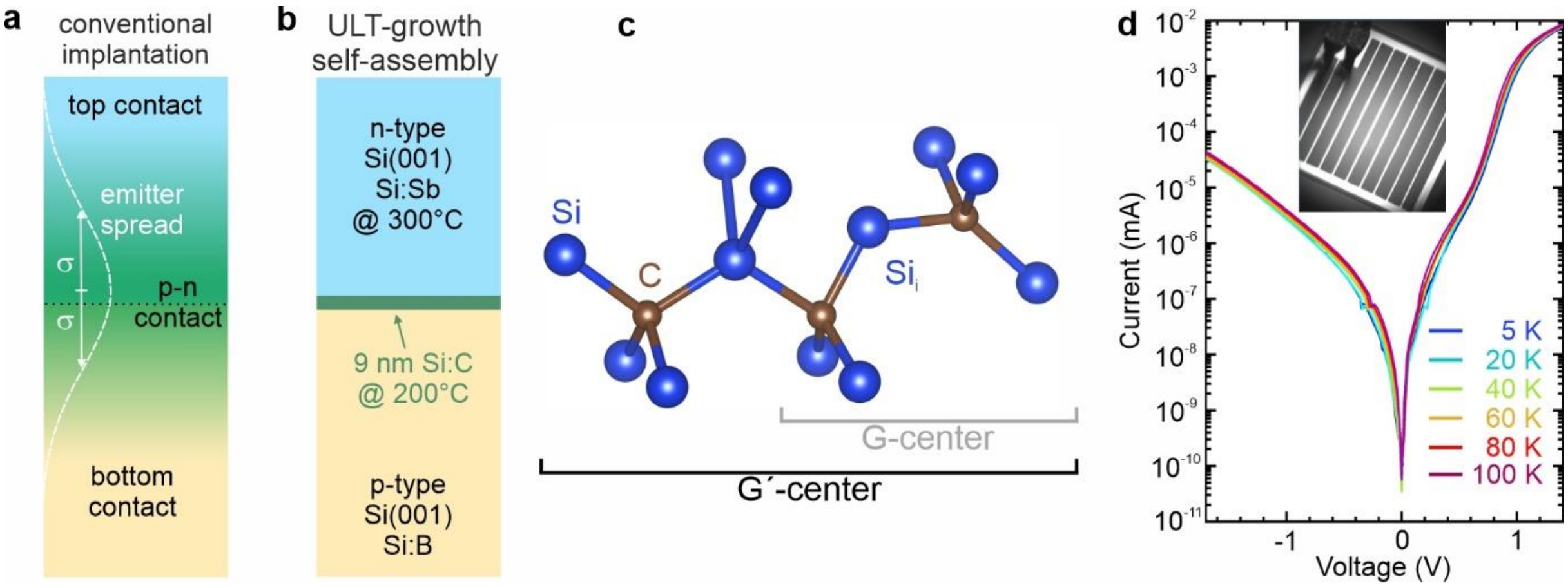

**Figure 1.** *Growth scheme and diode characteristics*. *(a) Typical fabrication scheme of group-IV-based color centers embedded in vertical p-n diodes. The implantation through the top contact leads to a wide emitter spread after implantation. (b) Si color center light emitting diode fabricated entirely without ion implantation. The SiCCs result from self-assembly during the growth of C-doped Si at ultra-low growth temperatures of 200°C. The n-type top contact was grown at a low enough growth temperature to avoid the dissociation of the SiCCs in the underlying thin nanolayer. (c) Atomistic model of the G'-center, consisting of a G-center and an adjacent substitutional C-atom. (d) Temperature-dependent I-V characteristics of the fabricated device, indicating diode behavior. The inset shows a photograph of the device under test.*

Figure 2(a) shows EL spectra, demonstrating emission from the SiCCs, confined to the 9 nm thick layer at the p-n junction. The blue spectrum was recorded for an applied excitation current density ($I_{Appl}$) of 625 mA/cm$^2$, and the red spectrum for $I_{Appl}$ = 1875 mA/cm$^2$. The spectral shape is dominated by the G'

ZPL at a wavelength of 1300.6 nm at 5 K that shifts to 1303 nm at 80 K. A pronounced phonon side band can be observed at longer wavelengths containing the local phonon mode (LPM) at 1406 nm at 5 K. Water absorption influences the spectral shape in the wavelength range of 1360 nm to 1420 nm.

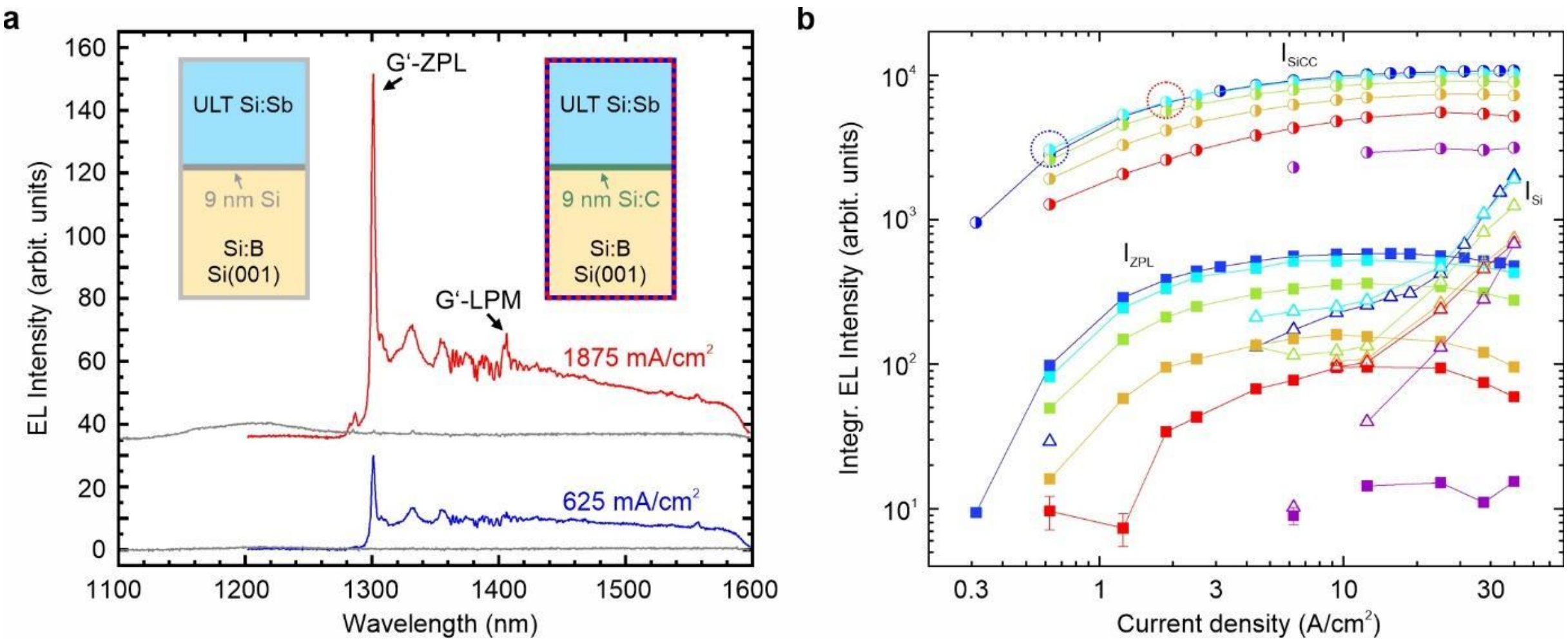

**Figure 2 (a)** *Electroluminescence spectra from a 9 nm thick SiCC layer obtained for applied current densities $I_{Appl}$ of 625 mA/cm$^2$ (blue) and 1875 mA/cm$^2$ (red). Gray spectra show the EL signal from the Si reference diode under the same excitation conditions. Spectra are vertically shifted for clarity. The inserts mark the respective sample layouts.* ***(b)*** *Sample temperature-dependent EL intensity versus current density for the total integrated color center emission $I_{SiCC}$ (half-spheres, 1270 nm - 1600 nm), the integrated intensity of the G' ZPL, $I_{ZPL}$ is indicated by full squares and the intensity from Si bulk by empty triangles. Sample temperatures of 5 K, 20 K, 40 K, 60 K, 80 K, and 100 K, are indicated by blue, cyan, green, orange, red, and violet color.*

The grey spectra in Fig. 2(a) mark the optical response from the pure Si diode obtained at the same applied current densities and for the same geometrical diode design. The insert in Fig. 2(a) depicts the sample schemes for the SiCC and Si reference diode, respectively. The clear difference in the integrated intensities indicates the efficient carrier capture and recombination in the SiCCs. Figure 2(b) presents the accumulated results of the integrated and temperature-dependent EL intensities of the SiCC sample versus excitation current density. Data points in blue, cyan, green, orange, red, and violet color correspond to sample temperatures of 5 K, 20 K, 40 K, 60 K, 80 K, and 100 K, respectively. The total integrated intensity of the SiCC emission, $I_{SiCC}$, is depicted by the half circles, the fitted integrated intensity of the G' ZPL is indicated by full squares, and the emission from bulk Si at higher current densities is plotted as open triangles. We note that already for very low current densities > 1 A/cm$^2$, saturation of the SiCC emission intensity is observed, indicating the low total amount of emission centers in the only 9 nm thick layer.

Figure 3(a) depicts temperature-dependent EL Spectra from the G'-center diode for an excitation current density of 1.88 A/cm$^2$ and sample temperatures of 5 K, 20 K, 40 K, and 80 K, indicated by blue, cyan, green, orange, and red color, respectively. This thermal quenching behavior of the G'-center EL is similar to that of conventional G-centers and is owed to the defect level structure within the Si band structure. Here, we found an activation energy for thermal quenching of the EL of the G'-center of

~25 meV. Figure 3(b) depicts excitation current density-dependent EL spectra obtained for a sample temperature of 5 K and $I_{Appl}$ ranging from 0.31 A/cm$^2$ to 4.38 A/cm$^2$. In the insert of Fig. 3(b), we plotted the integrated intensity of the ZPL of the G'-center emission ($I_{ZPL}$) versus $I_{Appl}$, which yields a superlinear function. The full-width at half maximum (FWHM) of the G'-ZPL is ~3 nm. Figure 3(c) aims to show that this relatively large FWHM is a direct consequence of the C-concentration used for creating the color centers in the Si:C layer. Notably, after an initial increase in the FWHM with $I_{Appl}$, the FWHM decreases with increasing $I_{Appl}$ from 3.6 nm to 3.2 nm, see inset of Fig. 3(b).

We explain these experimental findings by the nature of the G'-center in Si. In the intrinsic C-doped layer, the G'-center defect is in the optically active neutral state. When this ultrathin intrinsic layer is sandwiched into the p-n junction, many G'-centers will be charged due to the close proximity to the relatively thin depletion layer. By adding and increasing bias voltage into the forward region, the depletion region shrinks, and the quasi-Fermi level shifts toward the midgap at the edges of the ultrathin intrinsic layer. As a consequence, the G'-centers will be mostly neutral. This rise in the concentration of neutralized G'-centers goes exponentially with the applied bias voltage. By assuming that the Auger-ionization process is more efficient for the charged G'-centers than the neutral G'-centers, the carriers can recombine radiatively via the G'-center's deep states, causing electroluminescence. Thus, the superlinear $I_{ZPL}$ upon increasing bias voltage can be well explained by this effect. Indeed, the exponential function fits well with the observed data points in the inset of Figure 3(b). The ionization processes result in fluctuating charges around the electroluminescent G'-centers that cause spectral diffusion because of the Stark-shift in the G'-center emission, which finally broadens the observed FWHM in the ZPL emission. By stabilizing the charge state of G'-centers with increasing bias voltage, the concentration of charge fluctuators (charged G'-centers) around the electroluminescent G'-centers decreases, narrowing the FWHM in the ZPL emission of the electroluminescent G'-centers.

In the dotted lines in Fig. 3(c), we compare PL spectra of 9 mm thick SiCC layers grown at 200°C and overgrown by Si at a temperature of 300°C, for which different C-doping concentrations between $2.2 \cdot 10^{17}$ cm$^{-3}$ to $5.0 \cdot 10^{20}$ cm$^{-3}$ were used to create the G'-centers. With decreasing C-concentration in the nanolayer, a very pronounced linewidth narrowing of the G'-ZPL is evident from the spectral shape (Fig. 3(c). The grey data points in the inset of Fig. 3(c) depict the fitted FWHM of the G'-ZPL versus applied C-doping concentration, showing that the linewidth decreases from ~10 nm to less than 0.4 nm (< 300 µeV) for the lowest C-concentration. The red data point in the inset of Fig. 3(c) marks the FWHM obtained from EL, and the corresponding spectrum is plotted in red in Fig. 3(c). Thus, for the same applied C-doping concentration, the linewidths obtained from PL and EL are in excellent agreement, clearly pointing to the influence of the C-concentration on the FWHM of the G-center emission. We attribute this behavior to the non-uniform strain field produced by the neutral G'-centers themselves, which vanishes with the lower defect concentration.

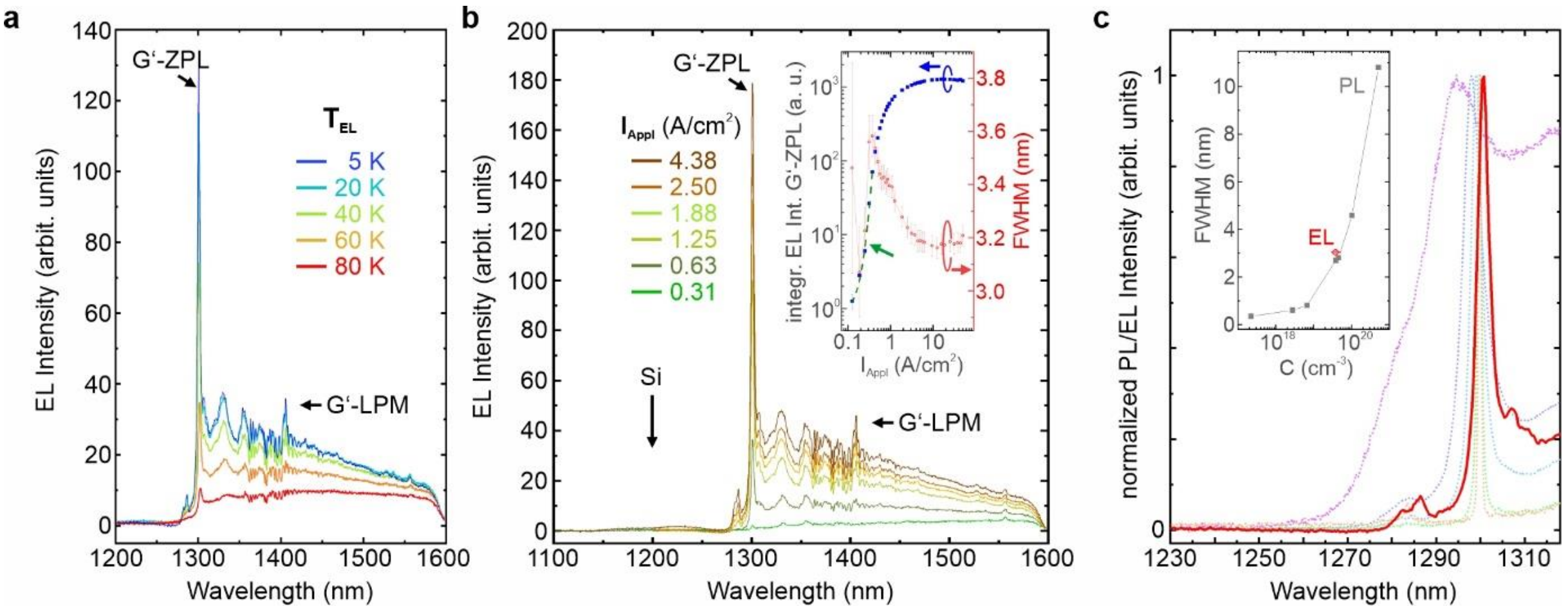

**Figure 3 (a)** *$T_{EL}$-dependent EL spectra from G'-centers embedded in a 9 nanolayer for an applied excitation current density $I_{Appl}$ of 1.88 A/cm$^2$. Blue, cyan, green, orange, and red color mark sample temperatures of 5 K, 20 K, 40 K, 60 K, and 80 K.* ***(b)*** *Current-density-dependent EL spectra for $I_{Appl}$ ranging from 0.31 A/cm$^2$ to 4.38 A/cm$^2$. The inset depicts the integrated EL intensity, $I_{ZPL}$ of the G'-centers versus $I_{Appl}$. The steepest slope shows a clear superlinear (exponential) increase of $I_{ZPL}$ versus $I_{Appl}$. The green dashed line shows an exponential fit to the data. The FWHM decreases as the in the superlinear regime.* ***(c)*** *Dotted spectra: PL response of G'-centers embedded in 9 nm thick layers and varying C-concentration, $2.2{\cdot}10^{17}$ cm$^{-3}$, $6.6{\cdot}10^{18}$ cm$^{-3}$, $4.5{\cdot}10^{19}$ cm$^{-3}$, $1.0{\cdot}10^{20}$ cm$^{-3}$, and $5.0{\cdot}10^{20}$ cm$^{-3}$, obtained at 5 K. The solid red spectrum shows the EL response of the SiCC layer at 5 K and for $I_{Appl}$ = 2.5 A/cm$^2$. The inset shows the change of the FWHM of the G'-ZPL with C-content. The red data point indicates that the linewidth was obtained from EL measurements.*

## Discussion

The fabrication of SiCCs without ion implantation is intimately linked to ultra-low temperature epitaxy, ULT, as presented here. In previous works, we found that Si, SiGe, and Ge layers, grown at ULT < 350°C, can be grown with high epitaxial quality if the chamber pressure during the growth is sufficiently low ($\lesssim 3\times10^{-10}$ mbar). In contrast to the growth at conventional, high temperatures >500°C, where the sticking coefficients of residual gas atoms on the substrate are small, at temperatures <350°C sticking coefficients become large, and atoms cannot be desorbed from the substrate surface [26], leading to their incorporation into the crystal. Figure S1 in the supplementary material presents the excellent maximum growth pressure of $3{\cdot}10^{-10}$ mbar, enabling good crystal quality despite low growth temperature. Such growth conditions also enable novel device schemes based on previously unattainable layer stacks of defect-free and fully strained SiGe and Ge heterostructure layers directly grown on Si or SOI [18,27,28]. Furthermore, type-I double heterostructure diodes can be achieved based on thick and Ge-rich SiGe/Si(001) layers [29].

Here, Si epitaxy at untypically low growth temperatures of 200°C concomitant with carbon co-doping enables a balance between crystalline quality, as seen from transmission electron microscopy [18], and kinetically-induced color center formation leading to the described G'-centers [21]. The results clearly demonstrate that an alternative fabrication path for SiCCs in diodes is feasible, notably without the use of ion implantation. The pure epitaxial nature of this fabrication scheme allows for the confinement of the SiCCs in layers of arbitrary thickness and at arbitrary vertical positions in the diode with nanometer resolution.

While the results of this work are a proof-of-concept structure, we note that for lasing applications, a large gain material volume, i.e., a high number of color centers, can be achieved through large layer thicknesses of tens to hundreds of nanometers, and high C-doping concentrations. Theoretical insights will be needed to optimize device and layer designs to find an optimum regarding, e.g., gain material volume, thickness of the width of the intrinsic layer of the p-i-n diode, and doping concentrations.

Additionally, we can envision the implementation of the self-assembled SiCCs into SiGe type-I double heterostructures that can result in an efficient carrier capture in the SiCC recombination region and limit minority carrier injection in the diodes [29]. Both, thick and fully strained SiGe layers for double heterostructure formation, and SiCC layers rely on epitaxy temperatures < 300°C. Thus, combining these two approaches can be considered to be possible without thermal deactivation and annihilation of the SiCCs. We note that silicon-on-insulator substrates (SOI) can be employed as well for the presented SiCC-diode growth scheme. This is particularly important for fabricating photonic resonators towards electrically-driven lasers and implementing isolated SiCCs for integrated quantum optics purposes.

While we have used a rather high C-concentration in this work to demonstrate the feasibility of the approach, we emphasize that SiCC self-assembly also opens the path to lower the SiCC density down to isolated emitter levels by either lowering the C-doping concentration, the layer thickness, or a combination of both accompanied by careful thermal annealing at appropriate temperatures. For quantum photonics applications, this all-MBE approach allows for embedding the SiCC quantum emitters in a matrix composed of isotopically purified $Si^{28}$.

## Conclusion

Color center-based LEDs suffer from a wide vertical spread of emitter position if the emitters are formed using ion implantation. This will ultimately limit, e.g., the usability of heterostructures to boost emitter properties. Here, we propose an alternative fabrication scheme for Si color center diodes and demonstrate that G'-centers, derivates of conventional G-center, can be embedded in an all epitaxial approach into a less than 10 nm thick layer that is deterministically grown at the p-n junction of the LED. Thereby, the emitters sustain the thermal budget needed for top contact formation and dopant activation. These results pave the way for applications of SiCCs as light emitters in Si photonics and single photon sources that can be manipulated through the electrical fields of the diode.

## Acknowledgements

This research was funded in whole or in part by the Austrian Science Fund (FWF) [10.55776/Y1238] and [10.55776/P36608]. For open access purposes, the author has applied a CC BY public copyright license to any author-accepted manuscript version arising from this submission. Support by the Ministry of Culture and Innovation and the National Research, Development and Innovation Office within the Quantum Information National Laboratory of Hungary (Grant No. 2022-2.1.1-NL-2022-00004) is much appreciated. AG acknowledges the European Commission for the projects QuMicro (Grant No. 101046911) and SPINUS (Grant No. 101135699).

**Supplementary Material**

# Telecom light-emitting diodes based on nanoconfined self-assembled silicon-based color centers

Andreas Salomon[1], Johannes Aberl[1], Enrique Prado Navarrete[1], Merve Karaman[1], Ádám Gali[2-4], Thomas Fromherz[1], Moritz Brehm[1]

## Growth parameters during molecular beam epitaxy

We argue that an excellent growth pressure is essential for ultra-low temperature growth. Impurity atoms that contribute to the background pressure and that are impinging on the sample surface during growth cannot be efficiently desorbed from the surface at temperatures lower than ~350°C [26]. Their incorporation into the crystal lattice leads to the formation of unwanted point defects and, eventually, at larger layer thicknesses, the breakdown of the epitaxial growth. Therefore, lower growth pressures during deposition significantly improve the layers' crystalline quality for ULT growth.

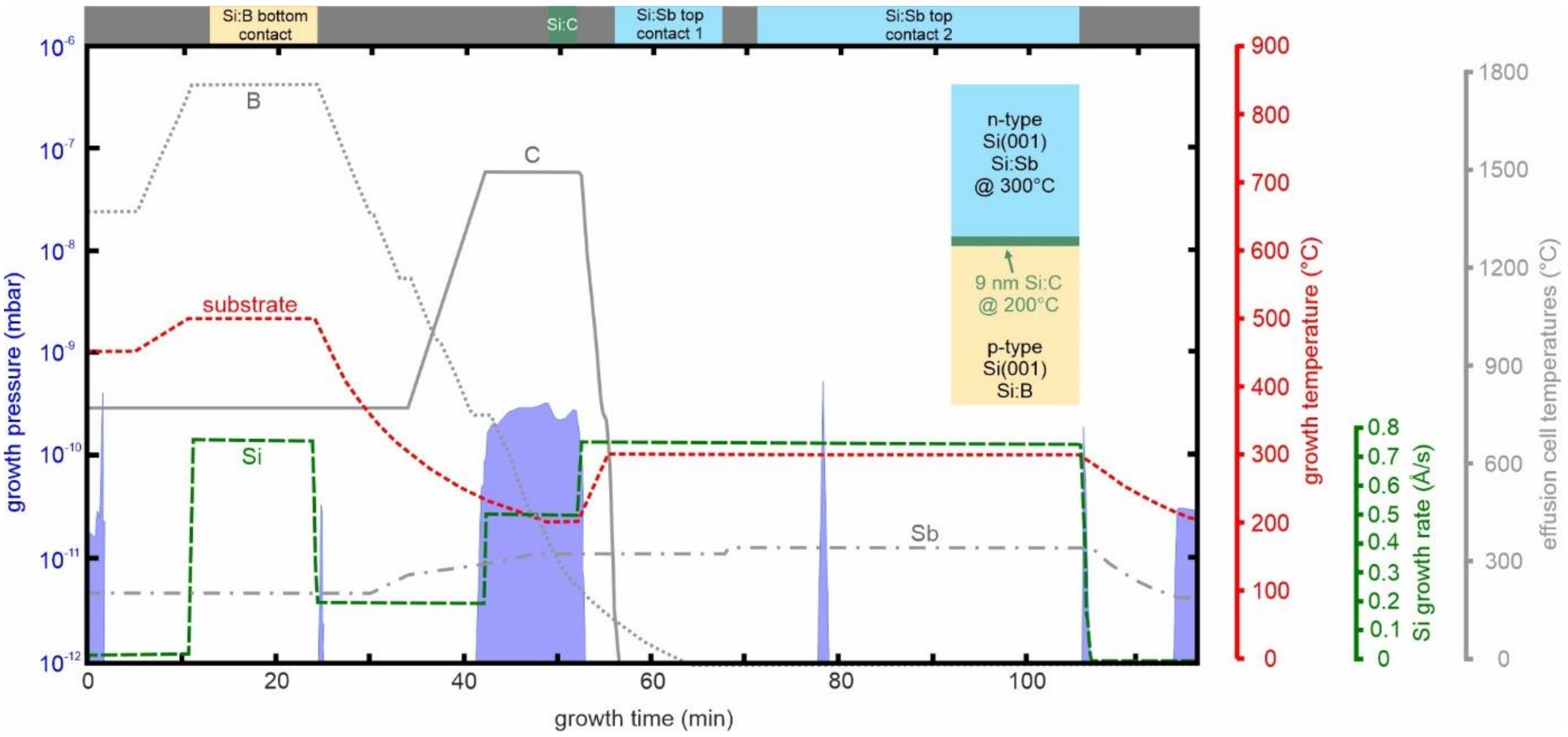

***Fig. S1:*** *Growth protocol of the LED containing self-assembled SiCCs. Very low growth pressures enable excellent layer qualities even for ultra-low-temperature growth. Growth pressure (left ordinate) is indicated by the blue-shaded areas. The right ordinates correlate to the substrate temperature (red dotted line), Si growth rate (green dashed line), and effusion cell temperatures for the carbon- (solid grey line), boron- (grey dotted line), and antimony source (grey dashed-dotted line). Within the grey-shaded areas, the growth temperature or the emission rates were ramped to their respective setpoints with all effusion cells and evaporators closed. The inset depicts the sample structure.*